# A Rule-based Model for Customized Risk Identification in Distributed Software Development Projects

Ansgar Lamersdorf
*University of Kaiserslautern*
*Kaiserslautern, Germany*
*a_lamers@informatik.uni-kl.de*

Jürgen Münch
*University of Kaiserslautern and Fraunhofer IESE*
*Kaiserslautern, Germany*
*Juergen.Muench@iese.fraunhofer.de*

Alicia Fernández-del Viso Torre
*Indra Software Labs*
*Madrid, Spain*
*afernandezde@indra.es*

Carlos Rebate Sánchez
*Indra Software Labs*
*Madrid, Spain*
*crebate@indra.es*

Markus Heinz
*University of Kaiserslautern*
*Kaiserslautern, Germany*
*m_heinz@informatik.uni-kl.de*

Dieter Rombach
*University of Kaiserslautern and Fraunhofer IESE*
*Kaiserslautern, Germany*
*Dieter.Rombach@iese.fraunhofer.de*

*Abstract*— **Many project risks in distributed software development are very different from the ones in collocated development and therefore are often overlooked. At the same time, they depend to a large extent on project-specific characteristics. This article presents a model for identifying risks early in a project. This model systematically captures experiences from past projects and is based on a set of logical rules describing how project characteristics influence typical risks in distributed development. Thus, the model is able to assess risks individually for each project. It was developed by applying qualitative content analysis to 19 interviews with practitioners. An evaluation using expert interviews showed that the risks identified by the model matched the actual experiences in 81% of the cases; of these, 40% have not been regarded yet at project start. The article describes the concepts of the model, its instantiation and evaluation, followed by a conclusion and future work.**

I. INTRODUCTION

The literature on distributed or global software development (GSD) is full of failure stories [1-3] caused by the inherent characteristics of GSD. Decreases in productivity [4, 5] or increases in the number of defects [6] have been reported as problems of GSD projects. Their causes include communication problems between sites [7-10], insufficient knowledge at one of the sites [6], mistrust between sites [11], or decreased workforce motivation due to the fear of job loss [12]. Therefore, they represent a set of GSD-specific project risks that might be relevant in addition to the typical risks of collocated projects.

This indicates that risk management in global development should specifically address the risks caused by the distributed nature of GSD projects. In practice, however, GSD-specific risks are often not considered at project start [13]. Instead, distributed projects are often initiated with a focus only on possible benefits such as low labor cost rates, while neglecting the problems of distributed development [14, 15].

On the one hand, knowing risks and potential problems together with their typical causes already at project start would help to initiate countermeasures and therefore reduce risks. On the other hand, this knowledge could also be used to systematically decide on the distribution of work to different development sites: If it is known which characteristics might cause certain problems (e.g., low expertise level, high turnover rate) and if these characteristics are known for the involved sites (e.g., the expertise level and turnover rate at each site), the decision on how to allocate work can take into account the possibility of specific problems at each site and weigh this against the potential benefits (e.g., a low labor cost rate).

In this article, we present a model for identifying and predicting GSD-specific project risks. It is based on a detailed qualitative content analysis of 19 interviews with practitioners regarding their experiences in distributed and global software development. From the interview analysis, we derived a set of rules that describe under which circumstances certain problems can occur. The rules use a set of influencing factors as independent variables that represent characteristics of the software development project environment. This allows for individually assessing the risks individually for a project based on project-specific characteristics and the distribution of work.

The remainder of the article is structured as follows. First, related work in risk identification for GSD is discussed. Section 3 presents an overview of the model concepts, followed by a detailed description of the interview study and the content analysis method used for model development in Section 4. In Section 5, the result of the interview analysis is presented as a set of rules and relevant characteristics. Finally, Section 6 explains the model evaluation within a Spanish software development company, followed by a discussion of the results and an outlook on future work.

## II. RELATED WORK

This section presents related work on risk identification for GSD. According to the Project Management Body of Knowledge (PMBOK) [16], risk management, a central aspect of project management, starts with risk identification followed by risk analysis. Risk identification addresses the question "Which risks might affect the project?", while risk analysis aims at evaluating the probability of occurrence and the severity for each risk. In the following, we will focus on the risk identification aspect of risk management.

There exists a large body of research on risk management and risk identification for software development projects [17-19]. However, these approaches usually do not consider distributed and global development. Thus, the specific risks of GSD are not adequately addressed in the general risk management literature. Consequently, we will concentrate on specific approaches for risk identification in GSD.

Prikladnicki et al. suggest a process for risk management that is integrated into processes for distributed software development [20, 21]. According to this approach, risk assessment is done on a tactical level during project planning and site selection. Based on the selection of sites, an individual document with offshore project risks is created that is then used for risk management during project execution. This approach demonstrates how risk identification and management can be done individually for every project and is interrelated with the selection of development sites. It mainly delivers a generic process without giving guidelines on how to identify the specific risks based on project and site characteristics. It can thus be seen as a generic process framework that needs to be filled with specific risk models for GSD.

Ralyte et al. present such a specific model for GSD, which includes a fixed set of risks that may occur due to the distributed nature of GSD projects [22]. Their risk framework is divided into the two dimensions distance (geographical, temporal, socio-cultural, organizational, technological, knowledge) and activity (communication, coordination, control, development, maintenance). For each combination of these two dimensions, they list specific problems that may occur in a project. The identification of these problems is based on a literature survey. In some cases, solutions for mitigating the risks are presented as well. For applying the framework to specific projects, the project-specific risks and solutions have to be selected from the proposed list.

A similar approach is given by Ebert et al. [23]. Here, several problems and risks that may occur in GSD projects are categorized into four drivers of global distribution: efficiency, presence, talent, and flexibility. The risks for each category are identified based on the authors' experience and a literature survey. While the number of identified risks is relatively small, this approach emphasizes the mitigation of risks and gives recommendations for overcoming the problems. Just like in the previous approach, this is mainly done on a generic level: The approach names a large number of possible problems and mitigation strategies and it is left to the user to identify which problems might occur in a specific project.

Smite [24] presents a risk identification approach that is more suited for identifying specific risks for an individual project situation. The approach distinguishes between threats, which in this case are possible negative situations (e.g., lack of experience, diversity in process maturity), and consequences to the project outcomes (e.g., budget overrun, time delays). Based on historical data, threats are linked to consequences together with a statistical probability of occurrence. It is thus possible to identify project-specific risks if the individual threats for each project are known. However, this approach relies on very detailed historical data and does not give an explanation on why a specific threat might lead to certain problems or consequences.

In general, current research on risk identification in GSD focuses very much on providing lists of possible problems and risks while giving no explanations or rules as to which problems might occur under which circumstances or in which environments. This, however, is very important in assessing project-specific risks, as significant project risks can depend on certain characteristics and constellations: Research shows, for example, that, depending on maturity, geographical distance between sites is seen very differently by project managers, from "no problem at all" to "a major barrier" [6], and that the consequences of staff turnover depend on the type of development project [25]. Therefore, there is a need for approaches in risk identification that consider the causal relations between project characteristics and problems and are able to assess risks individually for a specific project situation.

## III. MODEL GOAL AND BASIC CONCEPTS

Based on our previous work [25], we state the following two assumptions:

1) The specific problems and risks of distributed and global software development are often not known or underestimated at the beginning of GSD projects. 2) Most risks are not vital in all GSD projects but only under specific circumstances. Therefore, the following goal for the risk identification model was formulated:

*Goal: Develop a model that can be systematically used for the identification and assessment of risks in specific*

*global software development projects. The model should be based on previous experiences of practitioners in distributed and global development.*

As the goal is to identify project-specific risks, the model has to use the characteristics of a project environment as input for its predictions. Therefore, we decided to build the model as a set of rules stemming from interviews with practitioners in GSD and the experiences reported there. It can thus be seen as a formalized collection of lessons learned from previous projects.

The main elements of the model are (a) risks, (b) influencing factors, and (c) rules. Risks describe possible problems that might occur in a GSD project. Examples are "communication problems" and "lack of trust". For every risk, there exists a short textual description of its possible negative impact on the project (e.g., communication problems can decrease productivity; a lack of trust between teams can decrease both productivity and the motivation of the workforce).

Influencing factors describe characteristics of the project environment that have an impact on the existence of a certain problem. Influencing factors can be of different types: Characteristics of remote sites (e.g., the process maturity or the staff experience at the site), relationships between sites (e.g., the cultural difference or the existence of previous working experience between two sites), task characteristics (e.g., the complexity of a task), or characteristics of the overall project (e.g., the time pressure or the type of project). Task and site characteristics can be different for every involved site and might have to be elicited for each site. Relationships between sites have to be determined for every combination of two sites that collaborate in a project. Based on the experience of the authors, characteristics of the product to be developed or maintained might also be relevant. We model these characteristics indirectly as task characteristics.

In our model, we concentrate on software development within one organization; thus, we only look at the characteristics of different tasks. However, the model could also be used for evaluating risks in an outsourcing scenario. In such a case, the characteristics of the outside contractor should also be introduced as a set category of influencing factors.

Rules formalize how the influencing factors may have an impact on the risks. This can be done in two ways: While certain combinations of influencing factors can increase the possibility and severity of a risk, other combinations can decrease them. For example, one rule might be "cultural differences between sites increase communication problems", which describes the impact of one influencing factor on a risk. A more complex combination of influencing factors can be expressed in a rule such as "cultural differences and no previous working experiences between sites increase lack of trust". In this case, the risk is only affected if both influencing factors have a certain value. A decrease in the possibility and severity of a risk is defined, for example, by the rule "process maturity and previous experiences decreases communication problems".

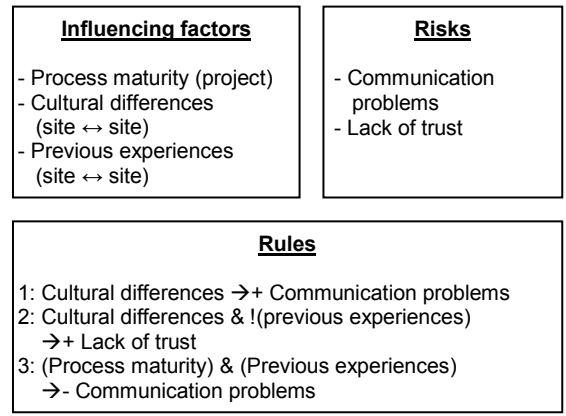

Figure 1. Exemplary model

The influencing factors in every rule can be combined using the logical operations ! ("not"), & ("and"), and | ("or"). Figure 1 shows a graphical illustration of these exemplary rules.

For assessing the risks of a global software development project, the model can be used as follows. Based on the estimations of the involved practitioners (e.g., project managers), the values for the influencing factors are assessed on a five-level scale (very low – very high). Project factors are assessed for the complete project, site and task factors for the remote sites, and relationships between sites individually for every two collaborating sites.

As a result of the value assessment of the influencing factors, the relevance of every rule can be evaluated (again on a five-level scale) according to the following recursive rules (in the example factor1: "high" and factor2: "low"):

1. If only one factor is on the left side of a rule, the relevance equals this value (e.g., "factor1 → risk X" has "high" relevance).
2. If the negation of one factor is on the left side, the relevance equals the negation of this value (e.g., "!factor1 → risk X" has "low" relevance).
3. If multiple factors are combined by an "and", the relevance equals the lowest value (e.g., "factor1 & factor2 → risk X has "low" relevance).
4. If multiple factors are combined by an "or", the relevance equals the highest value (e.g., "factor1 | factor2 → risk X has "high" relevance).

If the logical operators occur in more complex combinations, the rules can be applied recursively.

Figure 2 gives an example of the application of the rules. It can be seen that rule 2 has the lowest relevance while the relevance for rules 1 and 3 is high. The example also shows how risk assessment is dependent on work distribution: If the work was assigned to a different remote site with (for example) low cultural differences but also low previous experiences, the rules would be evaluated differently and other risks might have more relevance.

Based on the rule evaluation, the rules can then be ordered according to their relevance and the project-specific risks and problems can thereby be identified.

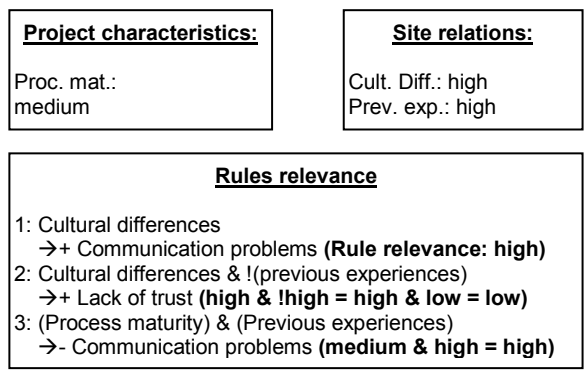

Figure 2. Assessment of influencing factors and evaluation of rule relevance for two sites

## IV. MODEL DEVELOPMENT

As a basis for the experiences captured in the model, we used a series of qualitative interviews with practitioners in global software development conducted between spring 2008 and fall 2009. Some of the interviews were conducted for a different study on task allocation practices in distributed development [25]. However, they also included questions on general experiences in distributed development and on factors causing problems in GSD.

In total, 19 interviews were conducted with experts from 14 different companies in the US, India, and Spain. The experts came from various different domains such as aerospace, educational software, and custom software development for the financial industry. With the exception of two interviewees who reported from a researcher perspective, all of them came from management positions, with 9 being project managers and others holding positions such as quality manager, product manager, or CIO. All interviewees could report from several (up to 20) years of experience in distributed and global software development projects.

While in most cases, only one interview was conducted per company, four interviewees came from Indra Software Labs (ISL), a software development division of a large Spanish multinational technology organization. ISL was later also used for evaluating the risk model (however, in a different interview session).

Each interview lasted for 30 – 75 minutes and was done (mostly) in person or over the telephone. With the exception of four interviews, all interviews were recorded and transcribed literally. For the other four, detailed notes were taken during and after the interview. This made it possible to analyze the interviews under various viewpoints.

According to the basic model, the interviews were analyzed with respect to statements on risks in distributed development, on factors influencing these risks, and on rules that describe experiences regarding how the factors impact the risks positively or negatively. This was done using qualitative analysis [26] and coding [27]: The code categories "Risk" and "Influencing factor" were created and all interviews were searched for codes that fit into these categories. Afterwards, the interviews were analyzed again and at where passages containing influencing factors and risks were identified, the experiences were extracted as rules combining the influencing factors and risks. Table 1 gives an example of how an interview passage is analyzed.

TABLE I. EXAMPLE OF TEXT ANALYSIS

| | |
|---|---|
| Original Passage | *"If you have a distributed team then it needs to be informed every time. If you have one single team, the management needs to inform only one team. […] the more sites, the more the number of teams that have to be coordinated"* |
| Identified Codes | Influencing factors:<br>• Number of sites<br>Risks:<br>• Coordination problems |
| Textual Rule | The more sites there are, the more people have to interact with each other in order to make any kind of decision and let the others know about it |
| Logical Rule | Number of sites →+ Coordination problem |

The first analysis of the interviews revealed a very large number of findings (42 influencing factors, 140 identified rules). Therefore, they were summarized: Some of the influencing factors and rules that described only singular experiences of specific organizations were left out. In addition, similar factors grouped together under one code and handled as one factor (e.g., the codes "infrastructure distance", "common tools", "availability of communication infrastructure", and "technological infrastructure" were grouped together). The same was done with the rules.

The summarization of the findings and the removal of some of the factors and rules were done based on the experience and evaluation of the authors and thus represent a threat to validity. However, as this was done following a defined process and documented throughout the process, the decisions were made transparently and can be traced back to the original findings in the interview transcriptions. As a result, we identified 31 influencing factors and 9 problems that are used in 46 rules formally describing the collected experiences on problem enablers and barriers in GSD.

## V. THE RISK IDENTIFICATION MODEL

In the following, a short overview of the identified model will be given. In order to make the model applicable in a specific environment, it was further customized and simplified: The model prototype was intended to be evaluated at ISL. Thus, influencing factors that were not relevant for ISL were removed from the model and rules that described the impact of these factors were also excluded. For example, the factor "outsourcing or captive offshoring" was removed, as all regarded development projects were done within ISL and did not include outsourcing to other companies. In addition, characteristics that could not be specified at project start (e.g., "capability to work independently") were also excluded. Finally, 23 factors and 36 rules were included in the model.

Table 2 shows all identified factors. They are categorized into relationships between the sites, characteristics of the site, characteristics of the task, and project characteristics.

An example of the identified rules was already given in Table 1. Most of the rules described problems at the interfaces between sites that impacted communication or coordination. The complete set of identified rules is presented in the appendix together with a textual description of each rule.

TABLE II. IDENTIFIED FACTORS

| Type | Factor | Explanation |
|---|---|---|
| Relationships between sites | Time zone difference | Differences between time zones at the sites |
| | Language difference | Differences in language or dialects in language (e.g., UK – India) |
| | Cultural difference | Differences in national or regional culture |
| | Personal relationships | Relationships between persons at different sites (have they met or talked personally?) |
| | Common working experiences | Experience of the two sites having worked together in the past |
| | Communication infrastructure | Quality of communication tools and network speed between sites |
| Characteristics of the site | Application knowledge | Expertise and knowledge in the application domain |
| | Technical knowledge | Expertise and knowledge in the technology (e.g., programming framework) |
| | Process knowledge | Knowledge about the development and communication processes used |
| | Transparency | Insight into remote site (plans, involved persons, status…) |
| | Staff motivation | Motivation of the staff for working on the project and in a distributed fashion |
| | Project experience | Experience of the personnel in similar projects |
| Characteristics of the tasks | Criticality | Criticality of the work (do failures threaten project success?) |
| | Complexity | Complexity of the work (e.g., needed documentation) |
| | Formality of the description | Degree of formality and fine granularity of task description |
| | Coupling to other tasks | Dependency and required communication between tasks |
| | Novelty of the product | Degree of novelty of the product for the involved persons |
| | Process phase | E.g., requirements, coding, testing |
| Project characteristics | Process maturity | E.g., CMMI level (can also be seen as site characteristic – in this environment, the sites in a project were of equal maturity) |
| | Product size | The size of the product to be developed |
| | Requirements stability | Degree of change in the requirements during the project |
| | Number of involved sites | Number of sites that need to collaborate |
| | Time pressure | Pressure on people working on the project |

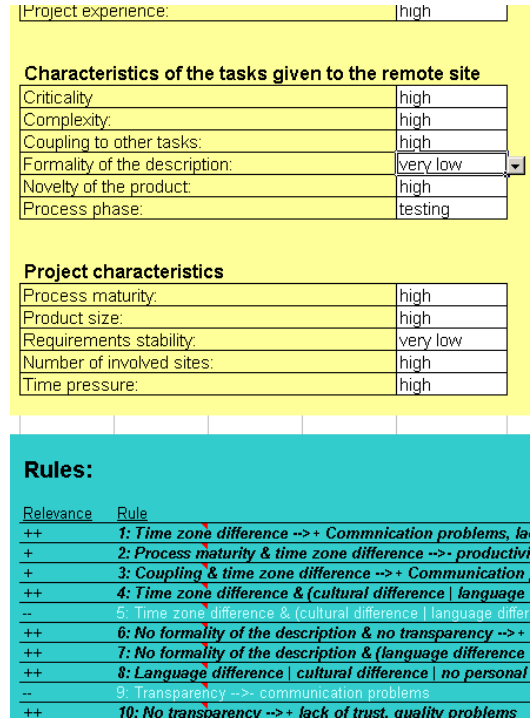

Figure 3. Screenshot of model implementation (excerpt)

Based on the identified factors and rules, the model was implemented in Microsoft Excel. Figure 3 shows a screenshot of the implemented model.

The implementation consists of two sections: In the project characterization part, the identified influencing factors can be set to project-specific values. In the rules section, all identified rules are shown together with their project-specific relevance. This relevance is automatically calculated based on the project characterization and the specification given in Section 3. For every rule, the textual description is included as a comment as shown in Figure 3.

Implementing the model in Excel made it possible to analyze a project and identify its individual risks and possible problems within minutes. This implementation was then also used in the evaluation of the model.

VI. EVALUATION

In the following section, we describe the evaluation of the prototype model with respect to the context and evaluation process, the results, and the threats to validity.

A. Context and Evaluation Process

The model was evaluated at Indra Software Labs (ISL) where four of the interviews for building the model were conducted (see Section 4).

Indra is the premier IT company in Spain and a leading IT multinational in Europe. ISL is the network of Software Labs of Indra that develops customized software solutions for Indra's markets. It has 20 development sites, half of which are located in Spain and the others in Latin America,

Slovakia, and the Philippines. Most of the software development projects at ISL are distributed either within Spain or globally. Therefore, within ISL there is a lot of experience with working in GSD projects and with related risks and problems.

We evaluated the model in interview sessions with five practitioners at ISL. Four of the interviews were conducted in person at an ISL site in Madrid, while the last one was done in a videoconference with the interviewee being located at Ciudad Real, Spain. The persons interviewed for the evaluation were different from the ones interviewed for the model development (see Section 4) and reported from different projects. There was thus no threat to validity, which might have arisen from using the same experiences for model development and evaluation.

Of the five interviewees, three were project managers, one was a director at ISL (responsible for one business area), and one was working in the quality department. From their perspective, they all had insights into various distributed projects in different constellations and had several years of experience. They were thus highly experienced in distributed software development.

The evaluation process was done as follows:
1. A questionnaire was sent to the interviewees in advance, asking them to recall one specific historical distributed development project. For this project, they were asked to characterize it and identify values for the 23 influencing factors.
2. In the interview, the model was used and all factors were set to the values named by the interviewees in the questionnaire. This resulted in an evaluation of the 36 rules with respect to the project characteristics.
3. Every rule that was identified as relevant (where relevance was wither "high" or "very high") by the model was presented to the practitioner (both as a logical rule and with the textual description) asking (a) whether the rule complied with the project experience (i.e., if its described impact on risks and problems could be observed) and – in the case that the rule did comply – (b) whether this rule was known at project start (i.e., if the project manager was aware of the phenomenon described by the rule).
4. Finally, the interviewees were asked if they found the use of such a model helpful and whether they would like to use it for future projects.

*B. Evaluation Results*

Table 3 shows the results of the evaluation. It shows that on average, one third of the 36 rules were relevant for each historical project. This indicates again that only a subset of the phenomena and problems of GSD described in the literature can be applied to a specific distributed development project.

A wide majority of the rules (81.4%) that were predicted as relevant could actually be observed in the projects. However, some rules were identified as irrelevant as they could not be observed in most of the projects: Rule 11 stated that a certain product size decreased the risk of losing intellectual property. This could not be confirmed by the practitioners because in their opinion, loss of intellectual property was never an issue at ISL, independent of project size. Rule 32 stated that if the coding phase was transferred to another site, project risks would be decreased, since coding tasks usually come with very detailed specifications. This could not be confirmed as the practitioners could report about various problems that also occurred when coding was transferred to another site. Despite these two rules, however, nearly all relevant rules could be confirmed in each of the five projects.

Of the rules that complied with the real project experience, nearly 59.5% were considered at project start by the project management. This means that the project managers were aware of a majority of the experiences stored in the model. One reason for this might be the fact that the interviews were conducted with highly experienced project managers who were aware of most of the risks and problems in distributed development and were able to incorporate these experiences into project planning. In less experienced environments, this number would therefore presumably have been lower.

However, a rate of 40.5 % still demonstrates that a significant proportion of the experiences stored in the model were not systematically regarded at project start. In these cases, an application of the model at project start would probably have helped, as it would have drawn attention to the described risks and made it possible to consider them in project management and to initiate countermeasures.

TABLE III. RESULTS OF EVALUATION

| Project No | # relevant rules | # rules confirmed (out of the relevant) | # rules considered at project start (out of the observed) |
|---|---|---|---|
| 1 | 14 | 12 | 8 |
| 2 | 16 | 12 | 5 out of 6 [1] |
| 3 | 10 | 9 | 7 |
| 4 | 9 | 6 | 2 |
| 5 | 10 | 9 | 3 |
| Σ | 59 | 48 | 25 out of 42 [1] |
|   |    | 81.4% | 59.5 % |

[1] The quality manager did not know for all rules if they were considered at project start or not

This hypothesis was also supported by the practitioners' answers to the applicability of the model: All of the five interviewed persons stated that they found the model useful and would like to use it in future projects. Even the managers who had already considered most of the experiences stored in the model (e.g., in projects 1 and 3) found the model very helpful: They reported that it was sometimes difficult for them to formulate their experiences and their predictions about possible risks and problems in meetings and discussions with other managers. In their opinion, such a model would help them demonstrate and communicate their experiences to others. Other managers stated that the model could also be used to identify and demonstrate project risks during project planning sessions with a customer.

Another advantage that was pointed out by one interviewed manager was the fact that this model can be used for evaluating different allocation scenarios: By inserting the characteristics of different remote sites into the model and assessing the predicted experiences and risks, the decision on how to select one out of different sites for a project could be supported.

*C. Threats to validity*

In the following, we will analyze the validity of the evaluation based on the four types of validity: internal (are the observed phenomena based on a cause-effect relationship?), conclusion (are the results statistically significant?), construct (do the measures reflect the real world?), and external (can the findings be generalized?):

A threat to the internal validity might be the fact that the interviewees did not understand the rules correctly while applying them for their projects. However, this threat was reduced by explaining every rule to the practitioners.

Conclusion validity is relatively low due to the small number of analyzed projects. To get a higher significance, a larger study should follow. However, the results seem to indicate a general trend as the degree of compliance (81%) is relatively high.

Construct validity might be threatened by the fact that the evaluation was conducted by the same person who developed the model and the interviewees might have been biased towards giving pleasant answers. Particularly, the question of whether the interviewees would like to use the model in later projects might have produced biased results. However, the significant rate of rules not considered at project start (40.5%) supports the usefulness of the model.

External validity might be threatened by the fact that all evaluation was done within one company. Therefore, it should be repeated at different organizations. However, as most of the interviews for model development (15 out of 19) were done in companies other than ISL, the evaluation results can probably be generalized.

VII.  CONCLUSION AND FUTURE WORK

In this article, we presented a model for assessing project-specific risks and problems of distributed software development. The model was described in its basic concept, its instantiation based on a systematic qualitative analysis of 19 interviews with practitioners, and its evaluation at Indra Software Labs.

The evaluation showed that the model was able to make predictions that complied with the experiences in historic projects and that its applicability in practice was strongly supported by highly experienced managers. This is probably due to the fact that, on the one hand, the model is relatively simple, which makes it easy to implement and easy to understand by practitioners. On the other hand, its underlying logical rules equip the model with some "intelligence" which results in customized rules and experiences on project risks that are relevant for a specific given project.

However, the model, as presented in this article still has some deficiencies that can be the basis for future work:

1) While the model already defines the concept of "influencing factors" relatively clearly and categorizes them into four groups, the risks and problems are not yet specified on a detailed basis. In the future, we would like to integrate the model into a larger model-based process for project planning, which would require a more formal definition of project risks and problems.

2) The current set of rules can be improved: While some rules seem to make predictions that do not comply with project experiences (see Section VI B), others are hard to understand, focus only on specific aspects, or overlap. This is mainly due to the fact that we tried to stick strictly to the results of the interview analysis, without abstracting too much from the transcribed statements or adding our own opinions. However, in the future, we want to establish a more systematic process for identifying rules for the model and for adding new rules. In this process, experienced practitioners would directly formulate their experiences as logical rules and add them to the model. After a project is finished, project managers could use their lessons learned to reformulate existing rules or add new ones. This would probably lead to an improved set of rules that would include more possible risks and problems of distributed development while at the same time creating less overlaps between rules.

If the model is used by software development organizations and regularly updated based on the experiences and lessons learned from finished GSD projects, it can be seen as a part of an Experience Factory [28]: The Experience Factory describes a way of systematic organizational learning, which includes experience modeling, experience storing, and experience reuse. Accordingly, the model presented here represents a way of modeling the experiences in distributed software development projects. The guidelines of the Experience Factory can also be used for updating the rules in the model: Based on the lessons learned in a new development project, the rule database is updated by increasing the confidence in a rule (if the rule correctly predicted the risks), changing a rule (if the rule did not predict the risks correctly), introducing new rules (if additional risks occurred in the project), or adding new influencing factors (if some rules were correct only in specific contexts, e.g., in certain types of projects).

In future work, we aim at integrating the model into a general process for project planning and site selection: As indicated by the practitioners, the model can be used for assessing different assignment scenarios with respect to the expected risks and problems. In addition, the influencing factors, problems, and causal relations can be reused as a basis for other models supporting task allocation decisions: In earlier publications, we developed models for selecting task assignments by using a model based on Bayesian networks [29, 30] or by using cost estimation models [31]. In both cases, we used the underlying concept that the impact of distributing work is dependent on certain influencing factors and causal relations. If these models are integrated into one coherent approach for systematic work distribution, the experiences stored in the risk assessment model can be used as input for the other models.

ACKNOWLEDGMENT

The authors would like to thank all participants in the interview and evaluation studies. At ISL, these persons were: Angel Villodre, Pablo Jesus Sanchez Moreno, Julian Diaz del Campo Jimenez de los Galanes, Francisco Fernández Fabián, Carlos Alger López, Diego Jiménez Romero, Ramón Torres, Ana López Díaz, David Graña, Ana Gómez-Escolar, Jose Luis Moragas, Manuel Cerrillo, Javier F. Gómez, Cindy Pinato, and Paloma Martínez Tordesillas. Some of the work was done during a stay at the Fraunhofer Center for Experimental Software Engineering, Maryland and was financially supported by the Otto A. Wipprecht Foundation. The authors also thank Sonnhild Namingha for proofreading the paper.

APPENDIX: IDENTIFIED RULES

| | Logical rule | Textual description |
|---|---|---|
| 1 | Time zone difference → +Communication problems, lack of trust | The bigger the time shift between sites, the less overlap exists between working hours. So there is less time to communicate or no time at all. Therefore, some of the product's problems might not be communicated sufficiently due to a lack of time.<br>If there is a higher time zone difference, people might need to wait longer for responses from the other site. A permanent delay might lead to a decrease in trust. |
| 2 | Process maturity & time zone difference → -Productivity drop | If there are very mature processes, it is possible to use the time shift for round-the-clock development; so more work can be done in less time. |
| 3 | Coupling & time zone difference → +Communication problems | If there is a time shift between two sites and there is only little time to communicate with each other, but there is a need for much communication because of highly coupled tasks, there is a problem. |
| 4 | Time zone difference & (cultural difference | language difference) → +Coordination problems | When there is a time zone difference there remains little time to communicate with each other. Having language and cultural differences in addition makes this even worse because the little time available can't be used efficiently when there is repeated mutual misunderstanding. |
| 5 | Time zone difference & (cultural difference | language difference) & (phase=requirements) → +Quality problems, risk of project failures | If the requirements phase is outsourced to another site that is far away from the customer in terms of language and culture, it is hard to get the requirements right in such a way that they reflect what the customer really wants. This increases project risks because later phases are based on the requirements. |
| 6 | !(Formality) & !(transparency) → +Risk of project failures | If it is not explicitly formulated what the other site has to do, they might not do things in the expected way. If there is no transparency, this might not be noticed early enough. |
| 7 | !(Formality) & (language difference | !(communication infrastructure)) → +Communication problems, productivity downfall | If the workforce doesn't understand exactly what to do and doesn't know who to ask or how to ask somebody, they can't do the work or have to wait for answers. |
| 8 | Language difference | cultural difference | !(personal relationships) | !(common experiences) → +Communication problems, lack of trust | If two persons can't get along communicating with each other, their trust in each other will suffer because each of them thinks that the other one isn't capable of understanding him due to a lack of competence. |
| 9 | Transparency → -Communication problems | If the persons at the other site and their schedule are known, people can communicate more efficiently and overcome communication problems. |
| 10 | !(Transparency) → +Lack of trust, quality problems | If there is no insight into what the other side is doing it isn't clear that they are doing anything at all. Therefore, there is some uncertainty as to whether deadlines can be met because of the missing parts that the other side should be working on. This causes mistrust. Additionally, the quality is reduced if people cannot coordinate with persons at the other site. |
| 11 | Size → +Travel cost overhead, -IP protection issues | The bigger a project is the more coordination work has to be done. This also means also more communication, which may include traveling to other sites or establishing more communication technologies.<br>If a project or an outsourced part is very large, it is very difficult to steal intellectual property without this being noticed. |
| 12 | Common experiences → -Productivity drop, coordination problems, lack of trust | If there is some experience of working together, one site knows how the people at the other site work and how they solve problems. Furthermore, there is not so much lack of trust: people know whom to talk to in case of problems and there is less "fear" of talking to them because they know each other. |
| 13 | Task coupling → +Productivity drop | Higher coupling means more communication is necessary and due to that, more time is spent on communication rather than on developing. |
| 14 | !(Process knowledge) & size → +Communication problems | A bigger project needs more communication and coordination. If there is a manager without experience in managing and coordinating a project correctly, there are a lot more problems in communication. |
| 15 | Language difference & cultural difference & !(common experiences) & !(personal relations) & !(process maturity) → +Risk of project failure | If there are differences in language, cultural, and work habits and there is no way to mitigate them using relationships or common experiences, they have to be handled by using a mature process. If there is no mature risk management strategy, then the whole project risk is much higher. |
| 16 | ((Cultural differences & !(maturity)) | time pressure) & !(project experience) → +Communication problems | If there are no experienced people, only immature processes, then occurring problems cannot be solved easily because people don't know how to solve them and have no guidance on how to do that. Cultural differences affect this in the way that people don't want to ask for help. If experienced staff is not available and (due to time pressure) cannot be acquired, the work can only be done more slowly or with lower quality. |
| 17 | Maturity & common experiences → -Lack of trust | Mature organizations tend to make reliable promises concerning quality, schedule, and budget. This - in combination with a common history in developing things where those promises were also kept - increases the level of trust. |
| 18 | !(Requirements stability) & (!(communication infrastructure) | !(maturity)) → +Coordination problems | If there are no stable requirements and a requirement changes, this change has to be communicated. This is not easily possible if there is no maturity or no good communication infrastructure between sites. |
| 19 | Process maturity → -Coordination problems, risk of project failures, cost overhead, quality problems, communication problems, lack of trust | Highly mature companies are more capable of dealing with problems and risks and even more efficient in doing that. |
| 20 | Application knowledge → -Productivity drop | The more familiar both parties are with the applications, the less knowledge needs to be transferred and the less communication is needed. Therefore, some time is saved. |
| 21 | Technical knowledge | application knowledge | process knowledge | personal relations → -Quality problems | If people on one site have no technical or application expertise or no experiences in working together, they make lots of mistakes. |
| 22 | Communication infrastructure → -Productivity problems, cost overhead, quality problems, communication problems | A good infrastructure makes communication easier. If both sites use the same tools and see the same times it is easier to communicate about something without many misunderstandings due to different visualizations. Therefore, productivity is higher. If there is a good communication infrastructure and it is used efficiently people don't have to travel that much for face-to-face meetings so that the overall cost is lower. |
| 23 | !(Communication infrastructure) & ( !(personal relations) | time zone difference) → +Communication problems | When there are bad tools and no personal relationships or a time zone difference, nobody wants to communicate because it costs too much time or is too difficult, and all that for the price of talking to a stranger who isn't trustworthy. |
| 24 | !(Communication infrastructure) & cultural difference → +Quality problem | Cultural differences and different habits of work have to be resolved partly with the help of common tools and a common infrastructure in order to avoid quality problems. |
| 25 | Staff motivation → -Quality problems | If people are involved in the distribution of tasks they are happier because they can work on the tasks they want to work on. So their motivation to work on that task is higher, which leads to better product quality. |
| 26 | !(Transparency) & time zone difference → +Productivity drop | If not much is known about a site, lack of trust can occur and productivity can decrease because people are not so willing to help the other site because of mistrust. The bigger the time difference, the less time to help them and reduce delays. |
| 27 | !(Transparency) & !(personal relationships) → +Risk of project failures | When one does not know the people at a site and their talents, experience, and capabilities, it is hard to be sure that they can actually do what they should do. Therefore there is a high risk for the project. |
| 28 | Transparency → -Lack of trust | The more is known about a site, the more it can be assessed how they are working and how well they are working. |
| 29 | Coupling & number of sites → +Communication problems | The more sites are involved in a project the more communication is necessary in order to coordinate work and solve occurring problems. This need is even stronger if the tasks of different sites are highly coupled. |
| 30 | Complexity | coupling → +Coordination problems | If you have complex and highly coupled task, it is hard to break them up into subtasks. Breaking them up causes coordination problems. |
| 31 | !(Cultural differences) & common experiences & communication infrastructure & process maturity → No problems | If there are no differences between sites and they are used to working together (in a high maturity process), there is nearly no difference to working in a collocated manner. |
| 32 | (Phase = coding) → -Project failure risk | If implementation is given to a remote site, typically complete specifications are given to the other site. Those are often easy to follow so the risk is lower. |
| 33 | (Phase = testing) & novelty of product & time zone difference & coupling → +Communication problems | Building entirely new products requires creativity and feedback from the users. This is very critical in the testing phase because that's a phase where the customer needs to be involved in order to recognize weaknesses of the product. Highly coupled tasks need more communication but this is hard to do with time shifts. So a lack of feedback can lead to a lot of mistakes and therefore a lot of rework to be done, which results in cost overhead. |
| 34 | Time pressure & !(personal relations) → +Communication problem | If people are under pressure they focus more on their work and are less willing to communicate. This is aggravated by a large distance and the lack of trust. So it is even more unlikely for them to communicate with the other site. |
| 35 | !(Requirements stability) & novelty of product & language difference & cultural difference → +Productivity downfall | If a new product is developed and requirements change, many things have to be discussed in order to go on. Language and cultural differences hamper these discussions. |
| 36 | Number of sites → +Coordination problem | The more sites there are, the more people have to interact with each other to make any kind of decision and let the others know about it; management structures have to be replicated. |